# The running coupling from the QCD Schrödinger functional – a one-loop analysis

Stefan Sint[a] and Rainer Sommer[b]

[a] Max-Planck-Institut für Physik
Föhringer Ring 6
D-80805 München, Germany

[b] CERN, Theory Division
CH-1211 Genève 23, Switzerland

**Abstract**

Starting from the Schrödinger functional, we give a non-perturbative definition of the running coupling constant in QCD. The spatial boundary conditions for the quark fields are chosen such that the massless Dirac operator in the classical background field has a large smallest eigenvalue. At one-loop order of perturbation theory, we determine the matching coefficient to the $\overline{\mathrm{MS}}$-scheme and discuss the quark mass effects in the $\beta$-function. To this order, we also compute the Symanzik improvement coefficient necessary to remove the $\mathrm{O}(a)$ lattice artefacts originating from the boundaries. For reasonable lattice resolutions and the standard Wilson action, lattice artefacts are found to be only weakly dependent on the lattice spacing $a$, while they vanish quickly with the improved action of Sheikholeslami and Wohlert.

## 1 Introduction

The strong coupling constant $\alpha_s$ can be extracted from experimental data, e.g. by comparing jet production rates at a collider experiment. On the other hand, several authors have emphasized that a theoretical determination of the strong coupling constant is possible by making use of lattice gauge theory [1–5]. This would provide a quantitative test of QCD which is believed to be the fundamental theory of both, hadronic physics and jet physics at high energies.

The basic strategy, as proposed by Lüscher, Weisz and Wolff [1] is easily described [6]. First, one fixes the free parameters in the QCD action by taking a corresponding number of low energy observables as experimental input. Then one has to find a suitable *non-perturbative* definition of a running coupling and trace its evolution from low to high energies, using Monte Carlo simulations. In the high energy regime, perturbation theory can be used to convert to other schemes such as the modified minimal subtraction scheme ($\overline{\mathrm{MS}}$) of dimensional regularization.

Since the above strategy applies to any asymptotically free field theory, it has first been tested for simpler models, namely the 2-dimensional non-linear O(3) model [1], and the pure SU(2) and SU(3) gauge theories [2–4]. One uses non-perturbatively defined couplings which run with $L$, the linear extension of the space-time volume. Then, using a recursive finite size scaling technique, the authors of refs. [1–4] were able to avoid the potential problem of having very different length scales fit on a single lattice. As a result, the running coupling was obtained deeply in the perturbative region, with all errors under control. The final perturbative matching to the coupling constant of the $\overline{\mathrm{MS}}$ scheme is done at the one-loop level in SU(3), and in the case of SU(2) even the two-loop coefficient is known [7].

Although the choice of the non-perturbative running coupling is not a question of principle, the practical feasibility of the project very much depends on it. Indeed, the great success in the pure gauge theories is largely due to a clever definition of the running coupling constant as the system's response to an external color electric field. The theoretical framework for such a definition goes under the name Schrödinger functional. By this one means the Euclidean path integral on a space-time manifold with boundaries at Euclidean times $x_0 = 0$ and $x_0 = L$, at which the values of the quantum fields are prescribed. The classical "path" then corresponds to a minimal action field configuration, the background field, which interpolates between the boundary values.

Before the Schrödinger functional can be used in practice, some theoretical considerations have to be made. In particular, the presence of the boundaries may lead to new singularities, which are not taken into account by the usual renormalizations of the bare parameters. Moreover, the lack of translation invariance on the space time manifold prevents the use of the standard proofs of perturbative renormalizability.

In the pure SU(N) gauge theory, these questions have been treated in ref. [8]. For a detailed account of the Schrödinger functional in QCD the reader should consult refs. [9,10]. Here we merely state the main result. Using power counting arguments, one expects [11] that the Schrödinger functional in QCD is renormalized by the usual QCD renormalizations of the coupling constant, the quark masses and, in addition, a multiplicative renormalization of the quark boundary fields. In particular, if the latter are taken to vanish, no additional divergence is introduced through the presence of the boundaries. This naive expectation has been confirmed at one-loop order of perturbation theory [10].

In this paper, we adapt the definition of the SU(3) running coupling constant to QCD and establish its perturbative relation to the $\overline{\rm MS}$ scheme at one-loop order of perturbation theory. The computation is straightforward in principle, but complicated through the presence of the quark masses. In particular, the dependence of the conversion coefficient and the $\beta$-function on the quark masses cannot be computed analytically and has to be extracted numerically from lattice perturbation theory. In view of the Monte Carlo simulations to be carried out later, we also determined the lattice artefacts in the observables.

The paper is organized as follows. In section 2, the definition of the Schrödinger functional is recalled both, in the continuum and on the lattice. The running coupling and the associated $\beta$-function are defined. In section 3, the conversion to the $\overline{\rm MS}$ scheme is done with and without the Sheikholeslami-Wohlert term in the action. There, we also discuss the $\beta$-function and the threshold effects in the perturbative running of the coupling. Lattice artefacts are discussed in section 4 and we finally summarize our main results.

## 2 The Schrödinger functional

### 2.1 Classical continuum action

A rigorous definition of the QCD Schrödinger functional exists on a space-time lattice and will be presented in section 2.6. On a more formal level, the Schrödinger functional can also be defined in the continuum [8,10]. It is given as the euclidean path integral with the action

$$S[A,\bar\psi,\psi] = -\frac{1}{2g_0^2}\int_0^L \mathrm{d}^4x\,\mathrm{tr}\,\{F_{\mu\nu}F_{\mu\nu}\} + \int_0^L \mathrm{d}^4x\,\bar\psi(\not{D}+m_0)\psi. \tag{2.1}$$

Here, $g_0$ denotes the bare coupling constant, $F_{\mu\nu}$ is the field tensor associated with the SU(3) gauge field $A_\mu$,

$$F_{\mu\nu} = \partial_\mu A_\nu - \partial_\nu A_\mu + [A_\mu, A_\nu], \tag{2.2}$$

and $D_\mu = \partial_\mu + A_\mu$ denotes the covariant derivative on the quark fields. For simplicity, we assume $n_f$ degenerate quark flavors of bare mass $m_0$, the generalization to the non-degenerate case being trivial. The $\gamma$-matrices are hermitian and satisfy

$$\{\gamma_\mu, \gamma_\nu\} = 2\delta_{\mu\nu}, \quad \mu,\nu = 0,\dots,3, \tag{2.3}$$

and the fifth $\gamma$-matrix is $\gamma_5 = \gamma_0\gamma_1\gamma_2\gamma_3$. Space-time is taken to be a cylinder of linear extension $L$. In the time direction, we impose the boundary conditions,

$$A_k|_{x_0=0} = C_k, \qquad A_k|_{x_0=L} = C_k', \qquad k = 1,2,3 \tag{2.4}$$

and

$$\begin{aligned} P_+\psi|_{x_0=0} &= 0, \qquad P_-\psi|_{x_0=L} = 0, \\ \bar\psi P_-|_{x_0=0} &= 0, \qquad \bar\psi P_+|_{x_0=L} = 0, \end{aligned} \tag{2.5}$$

with the projectors $P_\pm = \frac{1}{2}(1 \pm \gamma_0)$. In the spatial directions, periodic boundary conditions are imposed on the gauge fields, while the quark fields are taken to be periodic up to a phase factor,

$$\psi(x + L\hat{k}) = \mathrm{e}^{i\theta}\psi(x), \qquad \bar\psi(x + L\hat{k}) = \mathrm{e}^{-i\theta}\bar\psi(x). \tag{2.6}$$

Here, the spatial index $k$ takes the values $1, 2, 3$, and $\hat{k}$ denotes the unit vector in direction $k$. For reasons to be explained in section 2.7, we will later consider the values $0$ and $\pi/5$ for the parameter $\theta$.

## 2.2 The background field

The Schrödinger functional $\mathcal{Z}$ is considered a functional of the boundary gauge field $C$ and $C'$. In the following we restrict attention to the abelian boundary fields which have been introduced in ref. [3],

$$C_k = \frac{i}{L}\begin{pmatrix} \phi_1 & 0 & 0 \\ 0 & \phi_2 & 0 \\ 0 & 0 & \phi_3 \end{pmatrix}, \qquad C_k' = \frac{i}{L}\begin{pmatrix} \phi_1' & 0 & 0 \\ 0 & \phi_2' & 0 \\ 0 & 0 & \phi_3' \end{pmatrix}, \qquad k = 1,2,3, \tag{2.7}$$

with

$$\begin{aligned} \phi_1 &= \eta - \frac{\pi}{3}, & \phi_1' &= -\phi_1 - \frac{4\pi}{3}, \\ \phi_2 &= \eta(\nu - \frac{1}{2}), & \phi_2' &= -\phi_3 + \frac{2\pi}{3}, \\ \phi_3 &= -\eta(\nu + \frac{1}{2}) + \frac{\pi}{3}, & \phi_3' &= -\phi_2 + \frac{2\pi}{3}. \end{aligned} \tag{2.8}$$

This defines a 2-parameter family of boundary gauge fields. A solution to the field equations with these boundary fields is given by

$$B_0 = 0, \qquad B_k = \left[x_0 C_k' + (L - x_0)C_k\right]/L, \quad k = 1,2,3. \tag{2.9}$$

Moreover, for given boundary fields, $C$ and $C'$, the field $B$ represents the unique absolute minimum of the action. Any other field with the same action and boundary values is thus gauge equivalent to $B$, which will be referred to as the background field in the following. The associated field tensor $G_{\mu\nu}$ has the non-vanishing components

$$G_{0k} = \partial_0 B_k = (C_k' - C_k)/L, \quad k = 1,2,3, \tag{2.10}$$

which constitute the color-electric background field mentioned in the introduction.

### 2.3 The effective action

The uniqueness of the induced background field allows to unambiguously define the effective action of the Schrödinger functional as a functional of $B$, viz

$$\Gamma[B] = -\ln \mathcal{Z}[C', C]. \tag{2.11}$$

In ref. [10], the saddle point expansion of the Schrödinger functional has been carried out to one loop-order, using dimensional regularization. At this order one needs the fluctuation operators of the ghost, gluon and quark fields. For the precise definition of the pure gauge theory operators, $\Delta_0$ and $\Delta_1$, we refer to ref. [8].

Concerning the quark field fluctuation operator, it has been noted in ref. [9] that the Dirac operator

$$\mathcal{D} = \not{D} + m_0, \qquad D_\mu = \partial_\mu + B_\mu, \tag{2.12}$$

and its adjoint, $\mathcal{D}^\dagger$, allow for the definition of two distinct operators

$$\Delta_2 = \mathcal{D}\mathcal{D}^\dagger, \qquad \Delta_2' = \mathcal{D}^\dagger\mathcal{D}. \tag{2.13}$$

The operator $\Delta_2'$ is defined on spinors $\psi(x)$ which satisfy the boundary conditions (2.5). Furthermore, its eigenfunctions satisfy the modified Neumann conditions

$$(D_0 - m_0)P_-\psi(x)|_{x_0=0} = 0, \qquad (D_0 + m_0)P_+\psi(x)|_{x_0=L} = 0. \tag{2.14}$$

On the other hand, the operator $\Delta_2$ acts on fields which satisfy the same boundary conditions as $\bar{\psi}$ (2.5) and thus lives on a different space of functions. However, it follows from the analysis of ref. [9] that both operators, $\Delta_2$ and $\Delta_2'$, have exactly the same spectrum. For quantities which only refer to their spectrum we therefore need not distinguish between the two.

To write down the one-loop effective action, we make use of the $\zeta$-functions, defined through

$$\zeta(s|\Delta_i) = \mathrm{Tr}\,\Delta_i^{-s}, \qquad i = 0, 1, 2. \tag{2.15}$$

They extend to meromorphic functions in the whole complex plane, and one may show that their derivatives at $s = 0$ are well-defined. One then obtains

$$\begin{aligned}\Gamma[B] = &\left(\frac{1}{g^2_{\overline{\mathrm{MS}}}(\mu)} - \frac{33-2n_f}{48\pi^2}\ln\mu^2 - \frac{1}{16\pi^2}\right)\Gamma_0[B] \\ &- \frac{1}{2}\zeta'(0|\Delta_1) + \zeta'(0|\Delta_0) + \frac{1}{2}\zeta'(0|\Delta_2) + \mathrm{O}(g^2_{\overline{\mathrm{MS}}}),\end{aligned} \tag{2.16}$$

where $g_{\overline{\mathrm{MS}}}$ denotes the renormalized coupling constant in the modified minimal scheme ($\overline{\mathrm{MS}}$), and $\Gamma_0$ is the classical action of the induced background field,

$$\Gamma_0[B] = -\frac{1}{2}\int_0^L \mathrm{d}^4x\, \mathrm{tr}\,\{G_{\mu\nu}G_{\mu\nu}\}. \tag{2.17}$$

Note that the operators $\Delta_i$ have dimension $L^{-2}$. Therefore, the derivative of the $\zeta$-functions contains a logarithmic dependence on $L$, such that a logarithm of $\mu L$ is obtained in eq. (2.16). Furthermore it turns out that the field dependent part of the one-loop effective action is proportional to $\Gamma_0$.

## 2.4 Definition of the running coupling

As the effective action depends – apart from the quark masses – only on a single scale, $L$, eq. (2.16) suggests to define a renormalized coupling constant, $\bar{g}(L)$, through [8]

$$\left.\frac{\partial\Gamma}{\partial\eta}\right|_{\eta=\nu=0} = \frac{\left[\partial\Gamma_0/\partial\eta\right]_{\eta=0}}{\bar{g}^2(L)}, \qquad \left.\frac{\partial\Gamma_0}{\partial\eta}\right|_{\eta=0} = 12\pi. \tag{2.18}$$

Note that the derivative with respect to the parameter $\eta$ eliminates any divergent contributions to the effective action which do not depend on the background field.

Using the notation

$$\alpha_{\overline{\rm MS}}(q) = \frac{g^2_{\overline{\rm MS}}(q)}{4\pi}, \qquad \alpha(q) = \frac{\bar{g}^2(L)}{4\pi}, \quad q = 1/L \tag{2.19}$$

we are here interested in the perturbative relation

$$\alpha_{\overline{\rm MS}} = \alpha + c_1\alpha^2 + {\rm O}(\alpha^3)\ . \tag{2.20}$$

The coefficients in this expansion are functions of the parameter

$$z = \overline{m}(L)L, \tag{2.21}$$

where $\overline{m}$ is a suitably defined running quark mass. The form of these functions depends on the definition of $\overline{m}$. However, for the computation of the one-loop coefficient $c_1$ it is sufficient to define $\overline{m}$ at tree level. In the following we adopt the convention that, to this order, $\overline{m}$ coincides with the bare quark mass.

We split $c_1$ into its pure gauge theory value, $c_{1,0}$, and the quark contribution, $c_{1,1}$, viz

$$c_1 \equiv c_1(n_f, z) = c_{1,0} + n_f c_{1,1}(z), \tag{2.22}$$

where $c_{1,0}$ has been computed in ref. [3],

$$c_{1,0} = 1.25563(4). \tag{2.23}$$

A further renormalized quantity is obtained if the parameter $\nu$ in eq. (2.18) is kept different from zero, viz

$$\left.\frac{\partial\Gamma}{\partial\eta}\right|_{\eta=0} = 12\pi\left\{\frac{1}{\bar{g}^2} - \nu\bar{v}\right\}. \tag{2.24}$$

It is not difficult to see that $\bar{v}$ does not depend on $\nu$. Note also that $\bar{v}$ vanishes at tree level, because $\Gamma_0$ is independent of $\nu$. Its perturbative expansion reads

$$\bar{v} \equiv v_1(n_f, z) + \mathrm{O}(\alpha), \qquad v_1(n_f, z) = v_{1,0} + n_f v_{1,1}(z), \tag{2.25}$$

with the pure gauge theory contribution [3]

$$v_{1,0} = 0.0694603(1)\,. \tag{2.26}$$

The coefficients $c_1$ and $v_1$ can be calculated in the continuum through evaluation of eq. (2.16). One starts from the expression [8]

$$\frac{\partial}{\partial\eta}\zeta'(0|\Delta) = \lim_{\delta\to 0}\Big\{(\gamma_{\mathrm{E}} + \ln\delta)\frac{\partial}{\partial\eta}\alpha_0(\Delta) - \sum_{n=0}^{\infty} \mathrm{e}^{-\delta\lambda_n}\frac{\partial}{\partial\eta}\ln\lambda_n\Big\}, \tag{2.27}$$

where $\Delta$ stands for any of the fluctuation operators. The Seeley coefficient $\alpha_0$ is proportional to $\Gamma_0$ in all three cases, and $\lambda_n$ are the eigenvalues of $\Delta$, in ascending order.

The calculation of the quark contribution $c_{1,1}(z)$ essentially amounts to the determination of the eigenvalues of $\Delta_2$ and their derivatives with respect to $\eta$, up to a certain cutoff in the level $n$. One then evaluates the bracket in eq. (2.27) for a range of $\delta$-values and extrapolates to $\delta = 0$, taking into account that the bracket has an asymptotic expansion in powers of $\delta^{1/2}$ [8].

We employed a variational method with a plane wave basis to compute the eigenvalues. As a check on the precision, we also considered the eigenvalue equation for $\Delta_2$, which has a general solution in terms of hypergeometric functions. The boundary conditions eqs. (2.5),(2.14) then lead to a system of linear equations which can be numerically solved for the eigenvalues.

For $z = 0$, we were thus able to compute $c_{1,1}$ with an estimated numerical precision of 3–4 significant digits. As $z$ increases, cancellations in the sum over eigenvalues become stronger, resulting in a loss of precision. Since this computation was mainly intended to be a check on the more precise lattice methods, we do not quote the results here and refer the reader to section 3 instead.

### 2.5 The $\beta$-function

The Callan Symanzik $\beta$-function is defined through

$$\beta(\bar{g}) = -L\frac{\partial\bar{g}}{\partial L}\,, \tag{2.28}$$

where the derivative is taken at fixed bare parameters or, equivalently, keeping the renormalized parameters, $\overline{m}(L_0)$ and $\bar{g}(L_0)$, fixed at some normalization scale $L_0$. The $\beta$-function has a perturbation expansion

$$\beta(\bar{g}) \overset{\bar{g}\to 0}{\sim} -\bar{g}^3\sum_{n=0}^{\infty} b_n\bar{g}^{2n}\,, \tag{2.29}$$

with coefficients that are quark mass dependent,

$$b_n \equiv b_n(n_f, z), \qquad n = 0, 1, 2, \ldots \tag{2.30}$$

In particular, $b_0$ and $b_1$ coincide with the universal coefficients only when the quark mass is set to zero,

$$b_0(n_f, 0) = \frac{1}{(4\pi)^2}(11 - \frac{2}{3}n_f), \tag{2.31}$$

$$b_1(n_f, 0) = \frac{1}{(4\pi)^4}(102 - \frac{38}{3}n_f). \tag{2.32}$$

To obtain the mass dependence of the coefficients $b_n$, one may relate $\bar{g}$ perturbatively to any mass independent renormalized coupling constant. The coupling in the $\overline{\mathrm{MS}}$ scheme has this property. Writing

$$b_0(n_f, z) \equiv b_{0,0} + n_f b_{0,1}(z), \tag{2.33}$$

we obtain $b_{0,1}(z)$ from $c_{1,1}(z)$ through

$$b_{0,1}(z) = -\frac{1}{24\pi^2} - \frac{1}{8\pi} z c'_{1,1}(z), \qquad c'_{1,1}(z) \equiv \frac{\mathrm{d}}{\mathrm{d}z} c_{1,1}(z). \tag{2.34}$$

The central observable in a non-perturbative computation of the evolution of the coupling is an integrated version of the $\beta$-function, the step scaling function [1]. It is defined as follows. Starting with a value $u = \bar{g}^2(L)$ for the coupling at length scale $L$, the step scaling function, $\sigma$, is

$$\sigma(s, u, z) \equiv \bar{g}^2(sL)\,. \tag{2.35}$$

Finally, we note that a similar scaling function can be defined for the running quark mass.

## 2.6 Lattice formulation

### *2.6.1 The lattice action*

In Wilson's lattice QCD, the path integral representation of the Schrödinger functional reads [9],

$$\mathcal{Z} = \int \mathrm{D}[\psi]\mathrm{D}[\bar{\psi}]\mathrm{D}[U]\,\mathrm{e}^{-S}\,. \tag{2.36}$$

with the lattice action $S = S_g + S_f$, given by

$$S_g[U] = \frac{1}{g_0^2}\sum_p w(p)\,\mathrm{tr}\,\{1 - U(p)\}, \tag{2.37}$$

$$S_f[U, \bar{\psi}, \psi] = a^4 \sum_x \bar{\psi}(D + m_0)\psi. \tag{2.38}$$

The gauge field action $S_g$ is a sum over all oriented plaquettes $p$ on the lattice, with the weight factors $w(p)$, and the parallel transporters $U(p)$ around $p$. The weights $w(p)$ are 1 for plaquettes in the interior and

$$w(p) = \begin{cases} \frac{1}{2}c_s & \text{if } p \text{ is a spatial plaquette at } x_0 = 0 \text{ or } x_0 = L, \\ c_t & \text{if } p \text{ is timelike and attached to a boundary plane.} \end{cases} \tag{2.39}$$

The choice $c_s = c_t = 1$ corresponds to the standard Wilson plaquette action. However, these parameters can be tuned in order to reduce the lattice artefacts, as will be discussed in detail in section 4.

The Dirac operator in the quark action (2.38) is specified by

$$D = \frac{1}{2}\sum_{\mu=0}^{3}\{\gamma_\mu(\nabla_\mu^* + \nabla_\mu) - a\nabla_\mu^*\nabla_\mu\} + c_{\rm sw}\,\frac{ia}{4}\sum_{\mu,\nu=0}^{3}\sigma_{\mu\nu}P_{\mu\nu} \tag{2.40}$$

with the forward and backward covariant derivatives

$$\nabla_\mu\psi(x) = \frac{1}{a}[U(x,\mu)\psi(x+a\hat{\mu}) - \psi(x)], \tag{2.41}$$

$$\nabla_\mu^*\psi(x) = \frac{1}{a}[\psi(x) - U(x-a\hat{\mu},\mu)^\dagger\psi(x-a\hat{\mu})]. \tag{2.42}$$

The link variable $U(x,\mu)$ is the usual parallel transporter from point $x + a\hat{\mu}$ to point $x$, where $\hat{\mu}$ denotes the unit vector in $\mu$-direction.

The last term in eq. (2.40) has been introduced by Sheikholeslami and Wohlert [12], in order to cancel the leading cutoff effects of the standard Wilson quark action. It contains the lattice definition $P_{\mu\nu}$ of the field tensor,

$$\begin{aligned} P_{\mu\nu}(x) = \frac{1}{8a^2}\Big\{ & U(x,\mu)U(x+a\hat{\mu},\nu)U(x+a\hat{\nu},\mu)^\dagger U(x,\nu)^\dagger \\ & + U(x,\nu)U(x+a\hat{\nu}-a\hat{\mu},\mu)^\dagger U(x-a\hat{\mu},\nu)^\dagger U(x-a\hat{\mu},\mu) \\ & + U(x-a\hat{\mu},\mu)^\dagger U(x-a\hat{\mu}-a\hat{\nu},\nu)^\dagger U(x-a\hat{\mu}-a\hat{\nu},\mu)U(x-a\hat{\nu},\nu) \\ & + U(x-a\hat{\nu},\nu)^\dagger U(x-a\hat{\nu},\mu)U(x+a\hat{\mu}-a\hat{\nu},\nu)U(x,\mu)^\dagger\Big\} \\ & - (\mu \longleftrightarrow \nu), \end{aligned} \tag{2.43}$$

and our convention for $\sigma_{\mu\nu}$ reads

$$\sigma_{\mu\nu} = \frac{i}{2}[\gamma_\mu,\gamma_\nu]. \tag{2.44}$$

The standard Wilson quark action is recovered for $c_{\rm sw} = 0$, and the choice $c_{\rm sw} = 1$ will be referred to as Sheikholeslami-Wohlert action.

### *2.6.2 Boundary conditions and the background field*

The boundary conditions for the lattice gauge fields are

$$U(x,k)|_{x_0=0} = \exp(aC_k), \qquad U(x,k)|_{x_0=L} = \exp(aC'_k), \tag{2.45}$$

for $k = 1,2,3$, and with the abelian boundary fields $C$ and $C'$ as given in eq. (2.7). All other boundary conditions are as in the continuum (cf. section 2.1).

The boundary conditions (2.45) lead to a unique (up to gauge transformations) minimal action configuration $V$, the lattice background field. It can be expressed in terms of $B$ (2.9),

$$V(x,\mu) = \exp\left\{aB_\mu(x)\right\}, \tag{2.46}$$

and its field tensor (2.43) evaluates to

$$P_{\mu\nu}|_{U=V} = \frac{1}{a^2}\sinh a^2 G_{\mu\nu}, \tag{2.47}$$

with $G_{\mu\nu}$ given in eq. (2.10).

### *2.6.3 The effective action*

The effective action $\Gamma = -\ln \mathcal{Z}$ has an asymptotic expansion in the bare coupling constant,

$$\Gamma = g_0^{-2}\Gamma_0 + \Gamma_1 + \mathrm{O}(g_0^2) \tag{2.48}$$

with the lowest order term $\Gamma_0 = \left\{g_0^2 S_g[V]\right\}_{g_0=0}$. The next order term is, for $c_s = c_t = 1$, given by

$$\Gamma_1 = \frac{1}{2}\ln\det\Delta_1 - \ln\det\Delta_0 - \frac{1}{2}\ln\det\Delta_2\ . \tag{2.49}$$

Here, the operators $\Delta_i$, $i = 0,1,2$, are the lattice approximants of the continuum operators introduced above. Again, we refer to refs. [8,3] for the definition of $\Delta_0$ and $\Delta_1$. The operators $\Delta_2$ and $\Delta'_2$ are related to the lattice Dirac operator,

$$\Delta_2 = [(D+m_0)\gamma_5]^2, \qquad \Delta'_2 = [\gamma_5(D+m_0)]^2. \tag{2.50}$$

Since $D$ (2.40) acts on lattice spinors $\psi(x)$ which satisfy the boundary conditions (2.5), it follows that the eigenfunctions of $\gamma_5(D+m_0)$ (and thus of $\Delta'_2$) satisfy a lattice version of the modified Neumann conditions (2.14). Furthermore, from eq. (2.50) it is obvious that both lattice operators, $\Delta_2$ and $\Delta'_2$, have the same eigenvalues.

### 2.6.4 Definition of the running coupling

With these preliminaries, we define the running coupling constant $\bar{g}(L)$ and its relative $\bar{v}(L)$ through

$$\left.\frac{\partial\Gamma}{\partial\eta}\right|_{\eta=0} = k\left\{\frac{1}{\bar{g}^2} - \nu\bar{v}\right\}, \tag{2.51}$$

with

$$k = \left.\frac{\partial\Gamma_0}{\partial\eta}\right|_{\eta=0} = 12(L/a)^2[\sin(\gamma) + \sin(2\gamma)], \qquad \gamma = \frac{1}{3}\pi(a/L)^2. \tag{2.52}$$

Here, the normalization constant $k$ ensures that the renormalized coupling is exactly equal to $g_0$ at lowest order in the perturbation expansion.

## 2.7 Spatial boundary conditions and the Dirac operator

It remains to justify the choices 0 and $\pi/5$ for the angle $\theta$ in the spatial boundary conditions for the quark fields (2.5). Of course, setting $\theta = 0$ is a natural and aesthetically pleasing choice, which corresponds to periodic boundary conditions.

Beyond aesthetical criteria there is an additional, more technical one. Having in mind numerical simulations of QCD, it is worthwhile to recall that the speed of the known algorithms depends crucially on the condition number, i.e. the ratio between the highest and the lowest eigenvalue of the squared fermion matrix $\Delta_2$ (2.50). Thus it is desirable to have a lowest eigenvalue of $\Delta_2$, which is not too small.

| $\theta = 0$ | | | | $\theta = \pi/5$ | | | |
|---|---|---|---|---|---|---|---|
| $n$ | $\lambda_n$ | $n_c$ | $d$ | $n$ | $\lambda_n$ | $n_c$ | $d$ |
| 1 | 2.132449 | 2 | 2 | 1 | 4.693976 | 2 | 2 |
| 2 | 4.804360 | 2 | 2 | 2 | 4.881719 | 1 | 2 |
| 3 | 7.599922 | 3 | 2 | 3 | 8.384625 | 2 | 2 |
| 4 | 9.732686 | 1 | 2 | 4 | 13.109607 | 1 | 2 |
| 5 | 12.132366 | 3 | 2 | 5 | 13.834839 | 3 | 2 |
| 6 | 20.221614 | 1 | 2 | 6 | 19.717052 | 3 | 2 |
| 7 | 22.937850 | 2 | 2 | 7 | 26.100020 | 2 | 2 |
| 8 | 23.755510 | 2 | 2 | 8 | 26.996453 | 2 | 2 |
| 9 | 27.184931 | 1 | 6 | 9 | 27.826014 | 1 | 6 |
| 10 | 28.451394 | 3 | 6 | 10 | 27.846538 | 3 | 6 |

**Table 1:** The lowest eigenvalues (in units of $L^{-2}$) of the continuum operator $\Delta_2^c$ for vanishing mass and two choices of $\theta$. Since $\Delta_2^c$ is diagonal in color space, each eigenvalue is associated with a color component $n_c$. We also give the degeneracy $d$ for one quark flavor.

As a guiding principle, we have investigated the eigenvalues of the corresponding continuum operator[1] $\Delta_2^c$ (2.13). In ref.[9], it has been emphasized that the boundary conditions (2.5) introduce a gap, i.e. a non-vanishing lowest eigenvalue which is given by $\pi^2/4L^2 = 2.467.../L^2$, if the phase $\theta$, the quark mass and the background field are taken to vanish.

| | $L = 6a$ | | $L = 12a$ | | $L = 24a$ | | | |
|---|---|---|---|---|---|---|---|---|
| $n$ | $c_{sw} = 1$ | $c_{sw} = 0$ | $c_{sw} = 1$ | $c_{sw} = 0$ | $c_{sw} = 1$ | $c_{sw} = 0$ | $n_c$ | $d$ |
| $\theta = 0$ | | | | | | | | |
| 1 | 2.5338 | 3.0378 | 2.3147 | 2.5504 | 2.2190 | 2.3330 | 2 | 2 |
| 2 | 5.7261 | 5.0256 | 5.2428 | 4.9024 | 5.0181 | 4.8503 | 2 | 2 |
| 3 | 9.4614 | 10.4139 | 8.4086 | 8.8526 | 7.9758 | 8.1907 | 3 | 2 |
| 4 | 12.5904 | 14.8258 | 10.9562 | 11.9796 | 10.2964 | 10.7871 | 1 | 2 |
| 5 | 14.4516 | 13.3254 | 13.1233 | 12.5815 | 12.5924 | 12.3256 | 3 | 2 |
| 6 | 23.6179 | 20.7726 | 21.4968 | 20.1181 | 20.7701 | 20.0870 | 1 | 2 |
| 7 | 25.3644 | 26.1068 | 24.3912 | 25.1434 | 23.7033 | 24.0727 | 2 | 2 |
| 8 | 27.6740 | 26.9095 | 25.9626 | 25.1982 | 24.8980 | 24.5223 | 2 | 2 |
| 9 | 32.0603 | 33.9463 | 29.1583 | 29.7601 | 28.0767 | 28.2945 | 1 | 6 |
| 10 | 33.4272 | 34.4403 | 30.7130 | 31.0696 | 29.5507 | 29.6932 | 3 | 6 |
| $\theta = \pi/5$ | | | | | | | | |
| 1 | 5.7398 | 6.4868 | 5.1553 | 5.5042 | 4.9100 | 5.0788 | 2 | 2 |
| 2 | 6.0947 | 7.6777 | 5.4075 | 6.1333 | 5.1253 | 5.4732 | 1 | 2 |
| 3 | 9.9756 | 9.0438 | 9.0987 | 8.6484 | 8.7240 | 8.5022 | 2 | 2 |
| 4 | 15.2653 | 12.9957 | 14.0326 | 12.9231 | 13.5401 | 12.9897 | 1 | 2 |
| 5 | 17.6812 | 18.9750 | 15.4669 | 16.0670 | 14.5849 | 14.8749 | 3 | 2 |
| 6 | 23.6820 | 22.2339 | 21.2686 | 20.5772 | 20.4024 | 20.0625 | 3 | 2 |
| 7 | 29.9843 | 30.9829 | 28.0259 | 28.8318 | 27.0343 | 27.4287 | 2 | 2 |
| 8 | 32.6882 | 31.6622 | 29.8174 | 28.9978 | 28.3772 | 27.9760 | 2 | 2 |
| 9 | 31.9457 | 33.8957 | 29.7929 | 30.4575 | 28.7990 | 29.0545 | 1 | 6 |
| 10 | 33.8957 | 34.8946 | 30.2866 | 30.6210 | 28.9484 | 29.0768 | 3 | 6 |

**Table 2:** The lowest eigenvalues (in units of $L^{-2}$) of the lattice operator $\Delta_2$, for vanishing bare mass $m_0$. The eigenvalues are ordered according to their continuum limits (cf. table 1).

This picture remains valid in the presence of the abelian background field (2.9). At $\theta = 0$ and for vanishing quark mass, the lowest eigenvalue is slightly decreased and approximately given by $2.132/L^2$. Furthermore, it is possible to increase the lowest eigenvalue substantially by varying the angle $\theta$. We observed a maximal gap around $\theta = \pi/5$, and decided to consider this value besides $\theta = 0$. The lowest eigenvalues of $\Delta_2^c$ for these two choices are shown in table 1.

[1]In this subsection we add a superscript $c$ in order to distinguish the continuum operator from its counterpart on the lattice.

On a lattice with finite spacing $a$, the gap is somewhat larger. For illustration, we have collected the lowest eigenvalues of the lattice operator $\Delta_2$ on three lattices of different size $L/a$ in table 2. We observe that almost all low lying eigenvalues are approached from above by their lattice approximants, at a rate which is roughly given by $a/L$. In this respect, there is no difference whether the Sheikholeslami-Wohlert term is included in the action or not. We notice, however, that the approach to the continuum eigenvalues seems to be slightly more uniform for the Sheikholeslami-Wohlert action.

It is not clear how relevant these considerations are for a realistic simulation. Fluctuating gauge fields can lead to small eigenvalues of the lattice operator $\Delta_2$ and finally the typical spectrum has to be determined by numerical experiment. However, we believe that in sufficiently small volume, the dominant contributions to the path integral come from gauge field configurations which are close to the classical background field. Thus we suggest to use $\theta = \pi/5$ in practical applications.

## 3 One loop relations

The one-loop matching between $\alpha$ and $\alpha_{\overline{\rm MS}}$ (2.20) is done in two steps. First, we calculate the relation of $\alpha$ to a renormalized lattice coupling $\alpha_{lat}$. Combination with the known one-loop relation between $\alpha_{lat}$ and $\alpha_{\overline{\rm MS}}$ then yields the desired result.

### 3.1 The basic calculation

Insertion of the asymptotic expansion (2.48) into the definition of the running coupling, eq. (2.51), leads to the relation

$$\bar{g}^2 = g_0^2 + p_1\, g_0^4 + \mathrm{O}(g_0^6), \tag{3.1}$$

with $p_1$ being explicitly given by

$$p_1 = -\frac{1}{k}\frac{\partial \Gamma_1}{\partial \eta}\bigg|_{\eta=\nu=0}. \tag{3.2}$$

The coefficient $p_1$ depends on the number of flavors, the bare quark mass in lattice units, $m_0 a$, and the lattice size $l \equiv L/a$. For later convenience, this dependence is written in the form

$$p_1 \equiv p_1(n_f, z, l) = p_{1,0}(l) + n_f p_{1,1}(z, l), \tag{3.3}$$

where we have set

$$z = \overline{m}L, \qquad \overline{m} = \frac{1}{a}\ln(1 + m_0 a). \tag{3.4}$$

To this order, we thus identify $\overline{m}$ with the so-called pole mass, which is in one-to-one correspondence with the bare quark mass and coincides with the latter to leading order in the small $a$ expansion.

When the lattice spacing becomes small, the field dependent part of the one-loop effective action, $\Gamma_1$, is expected to diverge logarithmically. As usual, this divergence is absorbed in the renormalization of the coupling constant. We thus eliminate $g_0$ in favor of a renormalized coupling $g_{lat}$ (at the renormalization scale $\mu$), defined through minimal subtraction of logarithms [13], i.e.

$$g_0^2 = g_{lat}^2 + z_1(n_f, a\mu) g_{lat}^4 + \mathrm{O}(g_{lat}^6), \tag{3.5}$$

with

$$z_1(n_f, a\mu) \equiv z_{1,0}(a\mu) + n_f z_{1,1}(a\mu) = 2b_0(n_f, 0)\ln(a\mu). \tag{3.6}$$

Eq. (3.1) then becomes a relation between renormalized coupling constants,

$$\bar{g}^2 = g_{lat}^2 + (p_1 + z_1) g_{lat}^4 + \mathrm{O}(g_{lat}^6), \tag{3.7}$$

and we expect the coefficient in the r.h.s. of eq. (3.7) to have a well-defined continuum limit. Indeed, for $n_f = 0$, this has been demonstrated in refs. [8,3].

Therefore, we may restrict attention to the analysis of the quark field contribution, which, for $c_s = c_t = 1$, is given by

$$p_{1,1}(z, l) = (2kn_f)^{-1} \frac{\partial}{\partial \eta} \ln \det \Delta_2 \big|_{\eta=\nu=0}. \tag{3.8}$$

As we are considering spatially constant abelian background fields, eq. (2.9), $\Delta_2$ is diagonal with respect to its color and spatial momentum dependence. Its determinant factorizes accordingly and the problem is reduced to the evaluation of the determinant of a $4(l-1) \times 4(l-1)$ matrix in each subspace of fixed spatial momentum and color. In these subspaces, $\Delta_2$ acts as a second order difference operator in the Euclidean time, with coefficients that are $4 \times 4$ matrices in Dirac space. Following appendix C of ref. [8], the determinant of such an operator can be efficiently calculated by solving a recursion relation. Differentiation with respect to $\eta$ then leads to a recursion for the quantity of interest, i.e. the $\eta$–derivative of the determinant.

Alternatively, one may use the fact that $\Delta_2$ is the square of a first order operator (2.50). In the subspaces of fixed momentum and color, the determinant of $\gamma_5(D+m_0)$ can be computed by solving a first order recursion relation. We include a more detailed discussion of this method in the appendix.

We did the computation using either method and obtained $p_{1,1}(z, l)$ for lattice sizes up to $l = 64$ and in "REAL*16" precision. For fixed value of $z = \overline{m}L$, we expect $p_{1,1}$ to be given by an asymptotic series of the form [8]

$$p_{1,1} \underset{l\to\infty}{\sim} r_0 + s_0 \ln l + (r_1 + s_1 \ln l)/l + \mathrm{O}(\ln l / l^2). \tag{3.9}$$

The first few coefficients in eq. (3.9) can be extracted efficiently by first cancelling higher order terms in $1/l$ through numerical differentiation and then checking for stability as $l$ is increased [14]. Due to the large range of $l$ and the good numerical precision, this technique gives very accurate results for fixed values of $z$. In this way we determined the coefficient $s_0$ of the logarithm in eq. (3.9),

$$s_0 = -0.00844344(2). \tag{3.10}$$

This value has been obtained for $z = 0$, $\theta = \pi/5$ and $c_{\mathrm{sw}} = 1$. Furthermore, within the numerical accuracy, we found this result to be independent of all three parameters.

Recalling eqs.(3.6) and (3.7), we see that $s_0$ coincides with the expected result $s_0 = 2b_{0,1}(0) = -1/(12\pi^2) = -0.0084434319\ldots$. The continuum limit of the quark contribution to $p_1 + z_1$ thus exists and is given by

$$\lim_{a\to 0}[p_{1,1}(z, L/a) + z_{1,1}(a\mu)] = r_0(z) + s_0 \ln(\mu L). \tag{3.11}$$

We will set $\mu = L^{-1}$ in the following and assume the exact value $s_0 = -1/12\pi^2$ in the numerical analysis of the series (3.9). Since, in this section, we are interested in the universal relations between renormalized couplings in the continuum, we will concentrate on the conversion coefficient $r_0(z)$. The higher order terms in eq. (3.9) represent lattice artefacts and will be discussed separately in section 4.

### 3.2 Matching coefficient for massless quarks

The quark contribution to the matching coefficient between $\alpha$ and $\alpha_{lat} \equiv g_{lat}^2/4\pi$ is $4\pi r_0(z)$. For massless quarks, $r_0$ can be read off from table 3. Note that $\alpha_{lat}$ depends implicitly on the action, i.e. on the coefficient $c_{\mathrm{sw}}$.

| $\theta$ | $c_{\mathrm{sw}}$ | $r_0(0)$ |
|---|---|---|
| $\pi/5$ | 1 | $-0.034664940(4)$ |
| $\pi/5$ | 0 | $-0.009868186(4)$ |
| 0 | 1 | $-0.03328359(1)$ |
| 0 | 0 | $-0.00848683(1)$ |

**Table 3:** The first term in the asymptotic expansion (3.9) of $p_{1,1}$.

In order to obtain the one-loop relation between $\alpha$ and $\alpha_{\overline{\mathrm{MS}}}$, we also need the relation of $\alpha_{lat}$ to $\alpha_{\overline{\mathrm{MS}}}$,

$$\alpha_{lat} = \alpha_{\overline{\mathrm{MS}}} + d_1(n_f)\alpha_{\overline{\mathrm{MS}}}^2 + \mathrm{O}(\alpha_{\overline{\mathrm{MS}}}^3). \tag{3.12}$$

For the case of the Wilson action, the coefficient $d_1(n_f) \equiv d_{1,0} + n_f d_{1,1}$ is known in the literature [15,16]. In particular, the quark field contribution $d_{1,1}$ has first been calculated by Weisz [15], who gives the result in terms of the lattice integral $P_3$,

$$d_{1,1} = 4\pi P_3, \tag{3.13}$$

which was determined numerically with an accuracy of 3-4 significant decimal places. A more precise value for $P_3$ can be extracted from ref. [16], where a precision of 6 significant digits was achieved. We have re-evaluated $P_3$ and obtained

$$P_3 = 0.0066960(1), \tag{3.14}$$

which is, within errors, in agreement with the aforementioned values.

Taking $r_0(0)$ for $c_{\rm sw} = 0$ and both values of $\theta$ from table 3, we finally obtain the matching coefficients $c_{1,1}$ for massless quarks. They are both small and positive,

$$c_{1,1}(0) = -4\pi[P_3 + r_0(0)] = \begin{cases} 0.039863(2) & \text{for } \theta = \pi/5, \\ 0.022504(2) & \text{for } \theta = 0. \end{cases} \tag{3.15}$$

As already mentioned in section 2, these values are in agreement with the ones obtained from the direct expansion of the effective action in $\alpha_{\overline{\rm MS}}$.

### 3.3 Massive quarks

Let us now turn to the mass dependence of $\alpha$, which, at one-loop order, is described by $c_{1,1}(z)$, eq. (2.22). With our numerical methods we can compute $c_{1,1}(z)$ for fixed values of $z$, but these methods do not allow for a direct determination of the $z$-dependence. We therefore chose to represent the $z$-dependence of $c_{1,1}(z)$ by a Chebyshev approximation ($T_n(x) = \cos[n \arccos(x)]$)

$$c_{1,1}(z) - c_{1,1}(0) = z\left\{\sum_{n=0}^{11} t_n T_n(z/5 - 1) \ \pm 10^{-6}\right\}, \quad z \in [0, 10]. \tag{3.16}$$

The coefficients $t_n$ are listed in table 4. For most purposes, it is sufficient to use a truncated version of this Chebyshev approximation. For example, the first five terms describe $c_{1,1}(z) - c_{1,1}(0)$ with an accuracy of $5 \cdot 10^{-4} z$.

For large $z$, one needs a different approximation to $c_{1,1}(z)$. In order to arrive at an appropriate form, we note that a heavy fermion is expected to decouple from any physical prediction in the limit $z \to \infty$ [17]. An example is the relation between physical (and thus quark mass dependent) couplings. The fermionic contribution to the perturbative matching coefficients therefore should vanish for large values of $z$, up to power corrections in $1/z$. On the other hand, in unphysical definitions of the coupling, such as the $\overline{\rm MS}$ scheme, heavy quarks do not decouple. Therefore, we expect $c_{1,1}(z)$ to be logarithmically divergent in the large $z$ limit. Furthermore, the above argument shows, that not only the logarithm, but also a subleading constant contribution to $c_{1,1}$ would be the same in the relation between $\alpha_{\overline{\rm MS}}$ and any other physical coupling. In particular, in the case of the coupling defined via the static quark potential, this constant contribution is absent. We are therefore led to approximate

$$c_{1,1}(z) = -\frac{1}{3\pi}\ln(z) + \sum_{n=1}^{3} q_n z^{-n} \pm 5 \cdot 10^{-5}, \quad z^{-1} \le 0.13\,, \tag{3.17}$$

| $n$ | $t_n$ | | $w_n$ | |
|---|---|---|---|---|
| | $\theta = \pi/5$ | $\theta = 0$ | $\theta = \pi/5$ | $\theta = 0$ |
| 0 | −0.05996947 | −0.05560781 | −0.0066985 | −0.0066991 |
| 1 | 0.04146952 | 0.03840671 | 0.0057316 | 0.0084403 |
| 2 | −0.01440583 | −0.01451262 | −0.0017853 | −0.0045988 |
| 3 | 0.00411117 | 0.00535508 | 0.0002527 | 0.0020945 |
| 4 | −0.00068390 | −0.00187650 | 0.0001202 | −0.0007739 |
| 5 | −0.00013008 | 0.00062089 | −0.0001110 | 0.0002192 |
| 6 | 0.00016711 | −0.00019288 | 0.0000473 | −0.0000391 |
| 7 | −0.00007951 | 0.00005478 | −0.0000112 | 0 |
| 8 | 0.00002347 | −0.00001299 | | |
| 9 | −0.00000288 | 0.00000168 | | |
| 10 | −0.00000143 | 0.00000059 | | |
| 11 | 0.00000107 | −0.00000049 | | |

**Table 4:** Chebyshev coefficients of the fits eqs. (3.16),(3.20).

with

$$(q_1, q_2, q_3) = \begin{cases} (-0.10822, -0.0013, 0) & \text{for } \theta = \pi/5, \\ (-0.10718, -0.0334, 0.233) & \text{for } \theta = 0. \end{cases} \tag{3.18}$$

The coefficients $q_i$ $(i = 1, 2, 3)$, have been obtained by fitting eq. (3.17) to a number of data points $c_{1,1}(z_i)$ with $6 \leq z_i \leq 25$. To assess the quality of the fit, we varied the number of terms in eq. (3.17), and the number of points $z_i$ used for the fit. We conclude that $c_{1,1}(z) - c_{1,1}(0)$ is represented with a precision of $5 \cdot 10^{-5}$ for $z^{-1} \leq 0.13$.

### 3.4 Result for $\bar{v}$

The quantity $\bar{v}$ defined in eq. (2.51) is obtained in non-perturbative Monte Carlo simulations without any extra computational effort. Therefore it represents a useful check of the applicability domain of perturbation theory and the approach to the continuum limit [3].

In perturbation theory, we determined the contribution of massless quarks to eq. (2.25),

$$v_{1,1}(0) = \begin{cases} 0.0245370(1) & \text{for } \theta = \pi/5, \\ 0.013554(1) & \text{for } \theta = 0. \end{cases} \tag{3.19}$$

Again the quark mass dependence can be accurately described using a Chebyshev fit,

$$v_{1,1}(z) = \begin{cases} v_{1,1}(0) + z\ [\sum_{n=0}^{7} w_n T_n(z/5 - 1)\ \pm 5 \cdot 10^{-5}] & \text{for } z \in [0, 10], \\ 0\ \pm 5 \cdot 10^{-5} & \text{for } z > 8 \end{cases} \tag{3.20}$$

with the coefficients $w_n$ as given in table 4.

### 3.5 Λ-parameters

The one-loop coefficients between different renormalized couplings translate to ratios of the corresponding Λ-parameters. More precisely, if the one-loop relation between the couplings $g_a$ and $g_b$ reads

$$g_a^2 = g_b^2 + c_{ab} g_b^4 + \mathrm{O}(g_b^6), \tag{3.21}$$

the ratio between the Λ-parameters is given by

$$\Lambda_a/\Lambda_b = \mathrm{e}^{c_{ab}/2b_0}. \tag{3.22}$$

Here both, $b_0$ and $c_{ab}$ refer to the coefficients for vanishing quark masses. The reason is that Λ-parameters refer to the asymptotic high energy regime where all mass effects are negligible.

In the pure SU(3) gauge theory, the ratio between the Λ-parameters associated with the Schrödinger functional and the $\overline{\mathrm{MS}}$ scheme is

$$\Lambda/\Lambda_{\overline{\mathrm{MS}}} = 0.48811(1). \tag{3.23}$$

If the quark fields are included, this value is lowered. For instance, for $\theta = \pi/5$ and $n_f = 1$ ($n_f = 3$), the approximate value is 0.455 (0.383). Similar values are obtained with $\theta = 0$, and, recalling the relation between the Λ-parameters of the MS and $\overline{\mathrm{MS}}$ schemes,

$$\Lambda_{\mathrm{MS}}/\Lambda_{\overline{\mathrm{MS}}} = (\mathrm{e}^{\gamma_{\mathrm{E}}}/4\pi)^{1/2} = 0.37647475..., \tag{3.24}$$

we conclude that Λ is, for three or four quark flavors, almost equal to $\Lambda_{\mathrm{MS}}$.

As a byproduct of our computational strategy we also obtain the relation between the Λ-parameters $\Lambda_{SW}$ and $\Lambda_W$ , associated with the lattice couplings, $g_{lat}$, of the Sheikholeslami-Wohlert [12] and the standard Wilson action,

$$\begin{aligned} \ln(\Lambda_{SW}/\Lambda_W) &= n_f \left(r_0(0)|_{c_{\mathrm{sw}}=0} - r_0(0)|_{c_{\mathrm{sw}}=1}\right)/2b_0 \\ &= 0.024796754(2)\, n_f/2b_0. \end{aligned} \tag{3.25}$$

This relation holds for an arbitrary number of colors $N$ with $b_0 = (11N - 2n_f)/48\pi^2$. This is easy to understand when the matching of the couplings is done using the background field technique adapted to lattice perturbation theory [18,15,19]. Considering the 2-point function of the background field at one-loop order, the only difference between the two schemes are the quark loop diagrams with at least one quark-gluon vertex stemming from the Sheikholeslami-Wohlert term in eq. (2.40). Since this vertex is proportional to a group generator in the fundamental representation, the contraction of the group indices yields a $N$-independent constant.

As a further check of this statement, we performed the same computation using the abelian SU(2) background field of ref. [8] and obtained agreement within the numerical accuracy. Indeed, the number quoted in eq. (3.25) is the one obtained from this computation.

### 3.6 Threshold effects in the $\beta$-function

According to eq. (2.34), the mass dependence of the one-loop $\beta$-function can be obtained from $c_{1,1}(z)$. The derivative $zc'_{1,1}(z)$ which appears in $b_{0,1}(z)$ is, for large values of $z$, well approximated by taking the derivative of the fit eq. (3.17). On the other hand, for small values of $z$ we may differentiate the Chebychev fit, eq. (3.16). In this way, we obtained $zc'_{1,1}(z)$ with an estimated precision of $10^{-4}$ in the whole range of $z$-values.

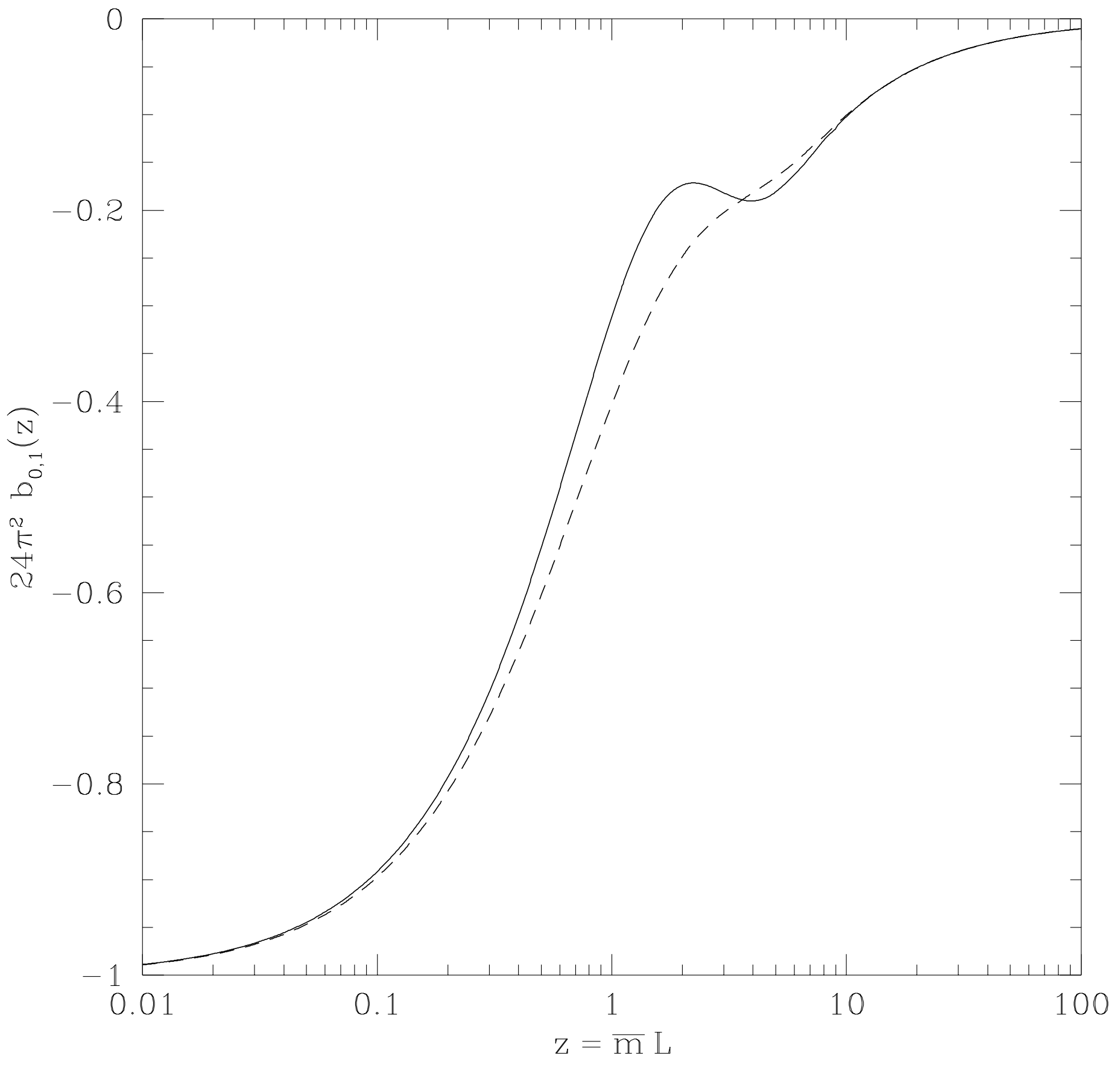

**Figure 1:** The contribution of one quark to the one-loop $\beta$–function, for $\theta = 0$ (dashed line) and $\theta = \pi/5$.

The resulting function $b_{0,1}(z)$ (2.33) is displayed in figure 1. One notices that the transition from an effectively massless quark to an approximately decoupled heavy quark is not very rapid. The reason is the following. $b_{0,1}(z) - b_{0,1}(0)$ starts with a term proportional to $z$ at small $z$. At the other end, the decoupling limit, $\lim_{z\to\infty} b_{0,1}(z) = 0$, is approached with a $1/z$-correction. These correction terms that are odd in the quark mass are possible because the boundary conditions eq. (2.5) do not respect chiral symmetry. This entails a rather broad transition region, in contrast to the case of other mass-dependent schemes, such as the $MOM$-scheme [20] or the coupling defined via the static quark potential.

At first sight, the existence of the $1/z$-term in the $\beta$-function of a heavy quark seems to pose a problem for the full non-perturbative computation of the $\beta$-function along the lines of ref. [3]. The reason is that quarks that are much heavier than $1/L$ contribute lattice artefacts that dominate over their physical effect if one is limited to lattice sizes of, say, $L/a < 20$. One would therefore like to omit quarks with $z > z_{\rm cut}$ from the Monte Carlo simulations, with a realistic $z_{\rm cut}$ taking values between, say, 2 and 4.

Our result allows to quantify the error that is made in the one-loop evolution of the coupling $\bar{g}$, if a quark with $z \geq z_{\rm cut}$ is omitted in the $\beta$-function. We fix the value of $\bar{g}(L_0)$ for some value $\overline{m}L_0 = z_0 \gg 1$ and want to know the coupling when $\overline{m}L = z \ll 1$. (One may think, e.g. of $L_0 = (0.5\,{\rm GeV})^{-1}$, $\overline{m} = 4.5\,{\rm GeV}$ and $L = (20\,{\rm GeV})^{-1}$). Omitting a heavy quark in the one-loop $\beta$-function corresponds to setting $b_{0,1}(z) = 0$ for $z > z_{\rm cut}$. The ensuing error is the difference in the couplings obtained by running once with the full and once with the truncated $\beta$-functions from scale $L_0$ to $L$. It is convenient to look at the corresponding error of $\bar{g}^{-2}$ which is given as the integral over the tail in the $\beta$-function between $z_{\rm cut}$ and $z_0$ (cf. fig. 1),

$$\Delta(\bar{g}^{-2}) = \frac{1}{4\pi}\Big\{c_{1,1}(z_0) - c_{1,1}(z_{\rm cut}) + \frac{1}{3\pi}\ln(z_0/z_{\rm cut})\Big\}. \tag{3.26}$$

Being generous, we enlarge the error by taking the limit of large $z_0$. This corresponds to taking the integral over the whole tail of $b_{0,1}(z)$, which approximately yields $\Delta(\bar{g}^{-2}) = 0.003$ for $z_{\rm cut} = 2$ and both $\theta = 0$ and $\theta = \pi/5$. In comparison, the present experimental error for $g_{\overline{\rm MS}}^{-2}(M_Z)$ from the LEP-experiments is about 0.05 [21], and the pure gauge coupling is known with precision $\Delta(\bar{g}^{-2}) = 0.02$ [3] (Note that the error of $\bar{g}^{-2}$ is preserved under one-loop evolution of the coupling).

If the real error is of the order of magnitude suggested by one-loop perturbation theory, one may conclude that heavy flavors with, say $z > 2$, can safely be omitted from a Monte Carlo simulation. On the other hand, perturbation theory suggests that quark flavors with $z \leq 2$ do not cause significant cutoff effects as we will see in section 4. Provided this picture is correct also beyond one-loop perturbation theory, the evolution of $\bar{g}$ can be computed through Monte Carlo simulations including the effect of a quark flavor of any mass and with all errors under control.

## 4 Lattice artefacts

The evaluation of finite lattice spacing effects in perturbation theory is important in several respects. First, we gain some insight about the cutoff effects to be expected in a full non-perturbative computation. Second, they are needed to determine the (Symanzik-) improvement coefficients and third, it has proven to be useful to define perturbatively improved observables [4], for which the one-loop cutoff effects are cancelled completely.

### 4.1 Symanzik's local effective Lagrangian

Consider a general lattice action for the Schrödinger functional. We expect that observables, such as $\bar{g}$, converge to their continuum values with dominant corrections that are linear in the lattice spacing $a$ (up to logarithmic corrections). This expectation is derived from the following reasoning. According to Symanzik [22] we can represent the lattice theory with finite spacing $a$ by a continuum theory, with a local effective Lagrangian containing the terms in eq. (2.1) plus higher dimensional operators accompanied by explicit powers of $a$. As long as these are not forbidden by an exact symmetry of the lattice action, we have to expect that all composite operators, which can be formed from the basic fields, are present in the local effective Lagrangian. In QCD with fermionic fields, gauge invariant dimension five operators exist. Integrated over the volume, they can contribute $\mathrm{O}(a)$ lattice artefacts [12]. In addition, in the Schrödinger functional, dimension four operators integrated over the surfaces at $x_0 = 0$ and $x_0 = L$ are a source of $\mathrm{O}(a)$ terms.

$\mathrm{O}(a)$ perturbative improvement eliminates these leading lattice artefacts order by order in perturbation theory. This is achieved by adding lattice representatives of the corresponding operators to the action one starts from. Their coefficients can then be tuned to cancel all $\mathrm{O}(a)$ effects to a given order in the coupling.

In the pure gauge theory, dimension five operators do not exist. Gauge invariance and axis permutation symmetry restrict the set of dimension four operators to $\sum_{j,k=1}^{3} \mathrm{tr}\{F_{jk}F_{jk}\}$ and $\sum_{k=1}^{3} \mathrm{tr}\{F_{0k}F_{0k}\}$. Our action (section 2.6) contains these operators with a strength tunable by the coefficients $c_s$ and $c_t$, respectively. For abelian background fields, as considered here, the first operator vanishes identically at the surfaces. Consequently, we do not obtain any information on $c_s$, here. In order to discuss perturbative improvement, we expand the second coefficient as

$$c_t = 1 + c_t^{(1)}(n_f)g_0^2 + \ldots\,, \qquad c_t^{(1)}(n_f) = c_t^{(1,0)} + n_f c_t^{(1,1)}\,. \tag{4.1}$$

Here, the $\mathrm{O}(1)$ term is fixed by requiring that the linear term is absent in the small $a$ expansion of $\Gamma_0$ and the pure gauge part,

$$c_t^{(1,0)} = -0.08900(5)\,, \tag{4.2}$$

is known from ref. [3].

Once we include the fermion fields, we have to consider two operators of dimension five [12]. Sheikholeslami and Wohlert have shown that one of these operators is redundant, when the theory posesses full translational invariance [12]. It is easy to see that the same is true for the Schrödinger functional. Consequently we include only the remaining operator, with its strength determined by $c_{\mathrm{sw}}$. Setting $c_{\mathrm{sw}} = 1$ corresponds to tree-level improvement. In the Schrödinger functional, one also needs to consider possible dimension four operators – such as $\partial_0(\bar{\psi}\psi)$ – at the boundaries. While we postpone a complete analysis of these terms to future work, we conjecture that they are included in our action with the correct weight for tree-level $\mathrm{O}(a)$ improvement. Indeed, this is indicated by the cutoff dependence of the observables $\bar{g}$ and $\bar{v}$, as will be seen below.

### 4.2 O($a$) improvement

Let us step back and discuss eq. (3.9) in more detail. Its structure is dictated by the aforementioned equivalence of the lattice theory with an effective continuum theory containing local interactions [22]. For example, the operators discussed above contribute artefacts that are linear in the lattice spacing. At one-loop order, the loop integration will generally introduce an additional logarithmic dependence on the lattice spacing.

These arguments lead us to expect that the continuum limit is approached in the form eq. (3.9) with the coefficients $r_0 \ldots$ being functions of $z$. For a tree-level improved action, $s_1$ should vanish. Indeed, setting $c_{\rm sw} = 1$, we find $s_1 = 0$ within our numerical accuracy of about $10^{-4}$ and for both values of $\theta$.

We are left with the linear term $r_1(z)/l$, which we determined numerically (for $c_{\rm sw} = 1$) to be

$$r_1(0) \quad = \quad -2c_t^{(1,1)} + \begin{cases} 0.038282(2) & \text{for } \theta = \pi/5, \\ 0.038282(2) & \text{for } \theta = 0, \\ 0.0382820(1) & \text{in the SU(2)-theory with } \theta = 0\ , \end{cases} \tag{4.3}$$

and

$$r_1(z) - r_1(0) \quad = \quad 0.012\ z\ \pm 2 \cdot 10^{-5} \quad \text{for } z \leq 10\ . \tag{4.4}$$

There are no local operators that could generate terms like $z^2/l$ or higher orders in $z$. This argument explains the structure of our numerical result (4.4). On the other hand, the linear term can be rewritten $z/l = am_0 + \mathrm{O}(l^{-2})$, corresponding to the operator $m_0 \mathrm{tr}\{F_{\mu\nu}F_{\mu\nu}\}$ in the local effective Lagrangian. In order to remove it in the improved action one just needs to perform a redefinition of the bare coupling $g_0$. For this reason, this term does not represent a genuine lattice artefact: its effect vanishes as soon as relations between renormalized quantities are considered. An example is provided by the step scaling function $\sigma$. Note that a similar reparametrization applied to the bare mass is expected to be necessary, when one considers the continuum limit at fixed $m_0 L$. This does not appear in our analysis, since we keep a physical mass fixed, when we take the continuum limit.

The only remaining linear lattice artefact is cancelled by chosing

$$c_t^{(1,1)} = 0.0191410(1)\,. \tag{4.5}$$

We conclude that the observable $\bar{g}^{-2}$ is O($a$) improved to one-loop order with our action (section 2.6) and the choices $c_{\rm sw} = 1$ and $c_t$ as quoted in eqs. (4.1),(4.2) and (4.5). Furthermore, a similar analysis for $\nu \neq 0$ shows that the same is true for the obervable $\bar{v}$, eq. (2.51). At present, we do not know, however, whether $S_f$ is the general tree-level improved fermion action, since the effect of two or more fermionic surface operators could in principle cancel each other in our observables. The general structure of the O($a$) improved action deserves further investigations.

### 4.3 Nonlinear lattice artefacts

We now discuss the complete one-loop lattice artefacts of the step scaling function. Besides the case $c_{\rm sw} = 1$ and $c_t^{(1,1)} = 0.019141$, we also consider the situation without improvement ($c_{\rm sw} = 0$ and $c_t^{(1,1)} = 0$). The latter may be of interest in Monte Carlo calculations because the Sheikholeslami-Wohlert term represents a significant numerical overhead in any simulation algorithm.

| | $c_{\rm sw} = 1$ and $c_t^{(1,1)} = 0.019141$ | | | $c_{\rm sw} = 0$ and $c_t^{(1,1)} = 0$ | |
|---|---|---|---|---|---|
| $L/a$ | $z = 0$ | $z = 1$ | $z = 2$ | $z = 0$ | $z = 2$ |
| 4 | 0.00009 | 0.00116 | 0.00197 | −0.00273 | 0.00035 |
| 5 | −0.00005 | 0.00069 | 0.00126 | −0.00330 | 0.00001 |
| 6 | −0.00010 | 0.00045 | 0.00088 | −0.00346 | −0.00020 |
| 7 | −0.00010 | 0.00032 | 0.00066 | −0.00344 | −0.00032 |
| 8 | −0.00008 | 0.00025 | 0.00051 | −0.00334 | −0.00041 |
| 9 | −0.00007 | 0.00020 | 0.00041 | −0.00322 | −0.00047 |
| 10 | −0.00006 | 0.00016 | 0.00033 | −0.00309 | −0.00051 |
| 11 | −0.00005 | 0.00013 | 0.00028 | −0.00296 | −0.00054 |
| 12 | −0.00004 | 0.00011 | 0.00023 | −0.00284 | −0.00057 |
| 13 | −0.00003 | 0.00009 | 0.00020 | −0.00273 | −0.00059 |
| 14 | −0.00003 | 0.00008 | 0.00017 | −0.00263 | −0.00060 |
| 15 | −0.00003 | 0.00007 | 0.00015 | −0.00253 | −0.00061 |
| 16 | −0.00002 | 0.00006 | 0.00013 | −0.00244 | −0.00061 |

**Table 5:** Lattice artefacts $\delta_{1,1}$ in the step scaling function for various $L/a$.

The lattice version of the step scaling function, defined exactly as in the continuum (cf. section 2.5) but for finite $a/L$, is denoted by $\Sigma(s, u, z, a/L)$. Its lattice artefacts have the perturbative expansion

$$\frac{\Sigma(2, u, z, a/L) - \sigma(2, u, z)}{\sigma(2, u, z)} = \delta_1(n_f, z, a/L)u + \mathrm{O}(u^2)\,, \tag{4.6}$$

with the one-loop coefficient

$$\delta_1(n_f, z, a/L) = \delta_{1,0}(a/L) + n_f\delta_{1,1}(z, a/L). \tag{4.7}$$

For the one-loop O($a$) improved action, the lattice artefacts in the pure gauge theory vary from $\delta_{1,0}(1/6) = -0.004$ to $\delta_{1,0}(1/16) = -0.0004$ (cf. table 1 in ref. [3]). Table 5 shows that even several massless flavors of quarks induce negligible lattice artefacts on this scale, when full $O(a)$ perturbative improvement is switched on.

In contrast, each massless flavor introduces $\delta_{1,1}(0, a/L) \approx -0.003$ lattice artefacts when we consider the action without improvement terms. Although the magnitude is not very large, there is a distinct problem. $\delta_{1,1}$ varies little with the lattice spacing in the accessible range of $a/L$. Only around $L/a = 16$ does the linear decay of the lattice artefacts set in, a behavior which is dominantly caused by the coefficient $s_1 \neq 0$ in eq. (3.9). Clearly, lattice artefacts of this type are difficult to detect in a simulation.

$\delta_{1,1}$ behaves similarly for values of $z$ up to $z = 2$. In particular, it decays roughly proportional to $(a/L)^2$, when improvement is switched on.

Very heavy flavors decouple in the continuum step scaling function up to $1/z$-corrections. When one stays at fixed $l$, however, and increases $z$, one enters the regime $am_0 > 1$. It is evident that one needs to avoid this region, where all fermionic effects are accompanied by large cutoff effects. One might think, however, that – as in the continuum – such a fermion decouples and it does not really matter whether one keeps it in the simulation or not. We want to briefly explain that there is a subtlety, here.

It is easy to expand $p_{1,1}$ for large $am_0$ (keeping $l$ fixed) in order to see qualitatively what happens in this limit. As long as $c_t^{(1,1)} = 0$, the fermion contribution to $\Sigma$ starts with a term $\propto 1/(1 + am_0) = \exp(-z/l)$. It decouples faster than in the continuum. When we switch on $O(a)$-improvement, however, a term $c_t^{(1,1)}/l$ remains, since a fermion with $am_0 > 1$ does not at all contribute the lattice artefact calculated in an expansion in $a$. Thus $O(a)$-improvement is ruined when $am_0 > 1$. Note that this behavior is not seen in table 5, since we restricted ourselves to $am_0 < 1/2$.

Let us finally point out that the exact numbers listed in table 5 depend on the convention that we chose for the renormalized quark mass, eq. (3.4). Nevertheless, we do not expect the qualitative behavior of $\delta_{1,1}$ to be very sensitive to the choice of $\overline{m}$. In particular, the large effect of tree-level improvement is caused by the vanishing of $s_1$ in eq. (3.9) with tree-level improvement. This is the case irrespective of the definition of $\overline{m}$.

## 5 Summary and conclusions

This article presents one step in the effort to compute the strong coupling from first principles following the original idea of refs. [1,8]. The challenge is to compute non-perturbatively the evolution of a renormalized coupling from low to high energies. As such a computation is to be performed through numerical simulations on finite systems, it is non-trivial to cover a large range of scales and take the continuum limit at the same time. If the coupling is defined in finite volume, running with the linear dimension, $L$, of the system, this potential problem can be solved by stepping up the energy ladder recursively. A suitable definition of a coupling has been known for the pure gauge theory [8,3].

Based on recent work on the QCD Schrödinger functional [9,10], we have extended the definition of the coupling to include quark fields in the functional integral. In particular we have made use of the freedom that exists in chosing the spatial boundary conditions for the quark fields. Requiring periodicity up to a phase, $\theta$, (cf. eq. (2.6)), this phase can be tuned such that the Dirac operator taken in the background field has a relatively large lowest eigenvalue. This occurs around $\theta = \pi/5$. In a numerical simulation in a sufficiently small volume, the gauge field fluctuates around the background field. Thus we expect that numerical simulations are eased by the corresponding choice of $\theta = \pi/5$.

We computed the renormalized coupling at one-loop order using both dimensional regularization and lattice perturbation theory. Our main result is the fermion contribution to the matching coefficient between the coupling in the Schrödinger functional and the $\overline{\mathrm{MS}}$ scheme. It is listed in eqs.(3.15–3.18) and table 4. We found that the quarks generally introduce rather small effects.

The Schrödinger functional provides us automatically with a mass-dependent coupling. When one crosses the threshold around $L^{-1} = \overline{m}$, the $\beta$-function changes rather slowly from its value for massless quarks to the large mass limit corresponding to the theory with that quark removed. One might be worried that such a slow threshold behavior is difficult to resolve in a Monte Carlo simulation, because quarks of a large mass introduce significant cutoff effects. Investigating this question quantitatively, we found, however, that possibly resulting uncertainties are small on the level of the statistical errors of corresponding pure gauge theory results. Thus there is no reason to expect that the threshold behavior seen at one-loop precision poses a problem in the numerical simulations.

As the final goal is to compute the evolution of the coupling in the continuum limit, it is important to investigate the size of lattice artefacts. They have been computed to one-loop accuracy for the step scaling function (2.35). We have considered two different lattice actions. I) The Wilson action, with the boundary terms taken exactly as in ref. [9]. II) The Wilson action plus the Sheikholeslami Wohlert improvement term removing $\mathrm{O}(a)$ "volume-effects" and the surface term that compensates for the operator $\sum_{k=1}^{3} \mathrm{tr}\{F_{0k}F_{0k}\}$ in Symanzik's local effective Lagrangian. The one-loop contribution to the coefficient of the latter is determined through the evaluation of the lattice artefacts that are linear in $a$. Its value is given in eq. (4.5).

The results obtained with the two actions are as follows. The overall size of lattice artefacts introduced by one quark flavor into the step scaling function is rather small, namely they are below the per-cent level for couplings $\bar{g}^2 < 3$ and reasonable lattices ($L/a \geq 4$). For the improved action (II) and massless quarks, they are even more than another order of magnitude smaller. Presumably the latter finding is rather accidental for our specific observable. A more important difference is the $a/L$-dependence of the lattice artefact for the Wilson action (I): at zero quark mass, it practically fakes a constant for all lattice sizes that are accessible to simulations. The slow vanishing of the lattice artefacts sets in only around $L/a = 16$.

Despite the fact that our explicit one-loop computations show that the step scaling function can be O($a$)-improved by including the two operators mentioned above, the complete structure of the O($a$)-improved action of the Schrödinger functional is not known, yet. It is possible that additional operators are necessary beyond one loop. Furthermore, even at the one-loop level we can not exclude that the improvement of other observables necessitates further operators in the action. This problem is currently investigated.

A further important question is a non-perturbative definition of the renormalized quark mass that can be well computed in the numerical simulations and shows small lattice artefacts. Also this problem is presently being addressed.

We would like to thank M. Lüscher, P. Weisz and U. Wolff for helpful discussions and critical comments at various stages of this work. One of the authors (S.S.) acknowledges financial support by CERN, where part of this work has been done.

## Appendix A

In this appendix we provide the details of our computation of $p_{1,1}(z,l)$ which makes use of a first order recursion relation for the lattice Dirac operator (cf. section 3). To increase the readability, we use lattice units ($a=1$, and thus $L=l$) and assume summation over repeated indices throughout the appendix.

The lattice operator $D_5 \equiv \gamma_5(D+m_0)$ (cf. eq. 2.40) is hermitian with real eigenvalues that come in pairs $\pm\sqrt{\lambda_n}$, with $\lambda_n$, ($n=0,1,\ldots,n_{\max}$) being the eigenvalues of $\Delta_2$ (cf. table 2 for the numerical values of the first few $\lambda_n$). Therefore, the determinant of $D_5$ is positive, and the quark field contribution $p_{1,1}(z,l)$ to the one-loop coefficient $p_1$ (3.1) can be written

$$p_{1,1}(z,l) = (k\,n_f)^{-1}\frac{\partial}{\partial\eta}\ln\det D_5\big|_{\eta=\nu=0}, \tag{A.1}$$

provided we have set $c_t=1$. Due to the special properties of the background field (cf. section 2), the eigenfunctions of the lattice operator $D_5$ are of the form

$$\psi(x) = \exp(ip_kx_k)\,u_{n_c}\,f(x_0). \tag{A.2}$$

Here, $\{u_{n_c},\, n_c=1,2,3\}$ denotes the canonical basis in color space,

$$u_1 = \begin{pmatrix}1\\0\\0\end{pmatrix}, \quad u_2 = \begin{pmatrix}0\\1\\0\end{pmatrix}, \quad u_3 = \begin{pmatrix}0\\0\\1\end{pmatrix}, \tag{A.3}$$

and $p_k$ are the spatial components of the allowed momenta

$$p_k = (2\pi n_k+\theta)/L, \quad k=1,2,3, \tag{A.4}$$

with integer numbers $n_k = 1, \ldots, L$.

According to the structure of the eigenfunctions, the determinant of $D_5$ factorizes and eq.(A.1) becomes

$$p_{1,1}(z,l) = \frac{1}{k} \sum_{n_c=1}^{3} \sum_{\mathbf{p}} \frac{\partial}{\partial \eta} \ln \det \mathcal{D}_5 \big|_{\eta=\nu=0}. \tag{A.5}$$

The reduced operator $\mathcal{D}_5$ acts on the functions $f(x_0)$ (A.2) which live in the subspace of fixed spatial momentum $\mathbf{p}$ and color $n_c$. Setting $x_0 \equiv t$, we have

$$(\mathcal{D}_5 f)(t) = -\gamma_5 P_- f(t+1) + \gamma_5 h(t) f(t) - \gamma_5 P_+ f(t-1), \tag{A.6}$$

with the boundary conditions (2.5)

$$P_+ f(0) = 0, \qquad P_- f(L) = 0. \tag{A.7}$$

To write down the explicit expression for the coefficient function $h(t)$, we introduce the notation

$$q_k(t) = \omega t + r_k, \quad \tilde{q}_k(t) = \sin q_k(t), \quad \hat{q}_k(t) = 2 \sin \frac{1}{2} q_k(t), \quad k = 1,2,3, \tag{A.8}$$

where $\omega$ and $r_k$ are related to the color component $n_c$ of the boundary fields (2.9),

$$\omega = (\phi'_{n_c} - \phi_{n_c})/L^2, \qquad r_k = p_k + \phi_{n_c}/L. \tag{A.9}$$

We then have

$$h(t) = 1 + m_0 + \frac{1}{2} \sum_{k=1}^{3} \hat{q}_k(t)^2 + i \tilde{q}_k(t) \gamma_k - \frac{1}{2} c_{\mathrm{sw}} \gamma_0 \gamma_k p_{0k}, \tag{A.10}$$

where $p_{0k}$ denotes the color component $n_c$ of the lattice field tensor (2.47) and further evaluates to $p_{0k} = i \sin \omega$, independently of $k = 1,2,3$.

At first sight, $\mathcal{D}_5$ seems to act like a second order difference operator. In fact, the projectors $P_\pm$ hide the first order structure which can be recovered through the following re-formulation. For $0 < t \le L$ we define a new function $F(t)$,

$$F(t) = P_- f(t) + P_+ f(t-1), \tag{A.11}$$

for which the boundary conditions (A.7) take the form

$$P_+ F(1) = 0, \qquad P_- F(L) = 0. \tag{A.12}$$

It is then straightforward to show that the eigenvalue equation, $(\mathcal{D}_5 - \mu) f = 0$, is equivalent to the recursion relation

$$F(t+1) = A(t) F(t), \tag{A.13}$$

with $A(t)$ being explicitly given by

$$\begin{aligned} A(t) \quad = \quad & -a(t)^{-1}\Big\{P_-\Big[\mu^2 - a(t)^2 + \mu\gamma_5(c_k(t)\gamma_k - b_k(t)\gamma_k + 1) \\ & \qquad\qquad + c_k(t)\gamma_k(b_j(t)\gamma_j - 1)\Big] \\ & \qquad + P_+\Big[b_k(t)\gamma_k - \mu\gamma_5 - 1\Big]\Big\}. \end{aligned} \tag{A.14}$$

Here, the coefficient functions are

$$\begin{aligned} a(t) &= 1 + m_0 + \frac{1}{2}\sum_{k=1}^{3}\hat{q}_k(t)^2, \\ b_k(t) &= i\tilde{q}_k(t) - \frac{1}{2}c_{\rm sw}p_{0k}, \\ c_k(t) &= i\tilde{q}_k(t) + \frac{1}{2}c_{\rm sw}p_{0k}, \qquad k = 1,2,3. \end{aligned} \tag{A.15}$$

If we now prescribe a value for $F(1)$, we may calculate F(L) for any $L > 1$ by solving the recursion relation (A.13),

$$F(L) = A(L-1)A(L-2)\cdots A(2)A(1)F(1). \tag{A.16}$$

At this point it is convenient to introduce the $2\times2$ matrix $M(\mu)$ in the subspace defined by the projector $P_-$, viz

$$M(\mu) = \big(B(L-1)\cdots B(1)\big)_{--}, \qquad B(t) = a(t)A(t). \tag{A.17}$$

The boundary conditions (A.12) for $F$ then correspond to the requirement

$$M(\mu)F(1)_- = 0, \tag{A.18}$$

with the non-vanishing 2-vector $F(1)_-$. To have a solution to eq.(A.18) the determinant of $M(\mu)$ must vanish. One then notices that $\det M(\mu)$ is a polynomial in $\mu$ of degree $4(L-1)$, and, using the same arguments as in appendix C of ref. [8], one concludes that $\det M(\mu)$ is proportional to the characteristic polynomial of $\mathcal{D}_5$. In fact, the correct normalization has been anticipated in eq. (A.17),

$$\det(\mathcal{D}_5 - \mu) = \det M(\mu), \tag{A.19}$$

so that one finally obtains

$$\frac{\partial}{\partial\eta}\ln\det\mathcal{D}_5 = {\rm Tr}\big(M^{-1}\frac{\partial}{\partial\eta}M\big), \qquad M \equiv M(0). \tag{A.20}$$

Note that the trace in the r.h.s. of this equation is over a $2\times2$ matrix. The inversion of the regular matrix $M$ is trivial and its derivative with respect to $\eta$ can be calculated using Leibniz' rule for the product of matrices in eq. (A.17).